\documentclass[twocolumn,aps,prl,groupedaddress,showpacs, superscriptaddress]{revtex4-2} 

\usepackage{amsmath}
\usepackage{amssymb}
\usepackage{txfonts}
\usepackage{pxfonts}
\usepackage{gensymb}
\usepackage{graphicx,bm,units,yfonts}
\usepackage[table]{xcolor}
\usepackage{hyperref}
\usepackage{movie15}
\usepackage{xcolor}

\DeclareUnicodeCharacter{2212}{-}

\begin{document}

\title{Isolating a Single Microtubule in Nanofluidic Device}

\author{Ssu-Ying Chen}
\email{sc945@njit.edu}
\affiliation{Department of Physics, New Jersey Institute of Technology, Newark, NJ, USA}

\author{Arooj Aslam}
\email{aaa249@njit.edu}
\affiliation{Department of Physics, New Jersey Institute of Technology, Newark, NJ, USA}

\author{David J. Apigo}
\email{david.j.apigo@njit.edu}
\affiliation{Department of Physics, New Jersey Institute of Technology, Newark, NJ, USA}

\author{Sagnik Basuray}
\email{sbasuray@njit.edu}
\affiliation{Department of Biomedical Engineering, New Jersey Institute of Technology, Newark, NJ, USA}

\author{Camelia Prodan}
\email{cprodan@njit.edu}
\affiliation{Department of Physics, New Jersey Institute of Technology, Newark, NJ, USA}

\date{\today}

\begin{abstract}
Biological systems have been theoretically predicted to support topological wave-modes, similar to the ones existing in meta-materials. The existing methods to measure these modes are not implementable to biological systems; new techniques have to be developed to accommodate measurements in life science. Motivated by this perspective, we report a nanofluidic device for studying one microtubule at a time. Micro-channels etched into fused-silica using reactive ion etching were interfaced with nanochannels milled with electron beam lithography, and sealed with a PDMS-coated glass coverslip. The microchannels are 1 $\mu$m deep and 100 $\mu$m wide, and the nanochannels are 150 nm deep and 750 nm wide, they are tested to be effective for isolating microtubules. The methods presented here are for an adaptable nanofluidic platform for phonon measurements  in biopolymers made of proteins or DNA.
\end{abstract}

\maketitle

\section{Introduction}

The field of topological mechanics and metamaterials is expanding to incorporate soft topological materials, such as micro and nano scale colloids and polymers, including biopolymers, suspended in water based media\cite{Prodan2009,Prodan2017} (CITE Vitelly). Phenomena similar to the one seen in topological insulators, like Thouless pumping, are theoretically proposed to exist in polymers suspended in fluids. In particular the the microtubules  and actin have been proposed to support topological phonon modes\cite{Prodan2009}, and has been inspiring the development of other lattices\cite{Prodan2017}. While the bending modes of the microtubules have been experimentally measured, the rest of the modes requires confinement and precise actuation.\cite{Aslam2019}. Previous experimental methods to obtain of the density of state of colloidal micro-particles in suspension require accurate  tracking of particle for a long period of time. (INSERT CITATION ISLAM colloids) To adapt such methods to microtubules, one needs to confine the microtubules to stay in the field of view of the microscope for a longer time, while allowing them to exibit dynamic instability.
The development of techniques in nanofabrication has become an increasingly useful tool in the life sciences. \cite{Kutchoukov2004,Iliescu2012,Cannon2004,Xia2008}  .\cite{Cao2002,Oz2019,Yeh2015}Methods that are developed for nanofluidic channels are adaptable for microscopy and  can integrate electronics for sensing and actuation of single proteins or macromolecules.\cite{Reisner2012,Kaji2004,Wang2005} Single proteins and macromolecules are typically imaged at high magnification using fluorescence or bright-field microscopy. This requires imaging through a thin optically transparent material, typically glass or fused-silica. Furthermore, the ability to adapt nanofluidic devices for electrodes to incorporate manipulation and sensing requires the nanochannels to be in an insulating material, and this is another benefit to fabricating in glass or fused-silica.

In this paper we present a method for confining a single microtubules in a nanochannel and a highly repeatable and detailed method for interfacing microchannels that are 50 $\mu$m apart with 750 nm wide by 150 nm deep nanochannels and capping them with a glass coverslip using PDMS as a bonding medium. The bonding is done outside the cleanroom and does not require expensive equipment. \cite{Haubert2006} The nanochannels are written into fused silica using electron beam lithography (E-beam lithography or EBL). This device can be used for transmission and fluorescence microscopy at high magnification with low fluorescence background. This method can be adapted to include electrodes in the nanochannels for sensing and actuation with dimensions that promote high signal to noise ratio. 

\section{Fabrication of Nanofluidic Devices in Fused Silica}

\begin{figure*}[!ht]
\centering
\includegraphics[width=0.8\linewidth]{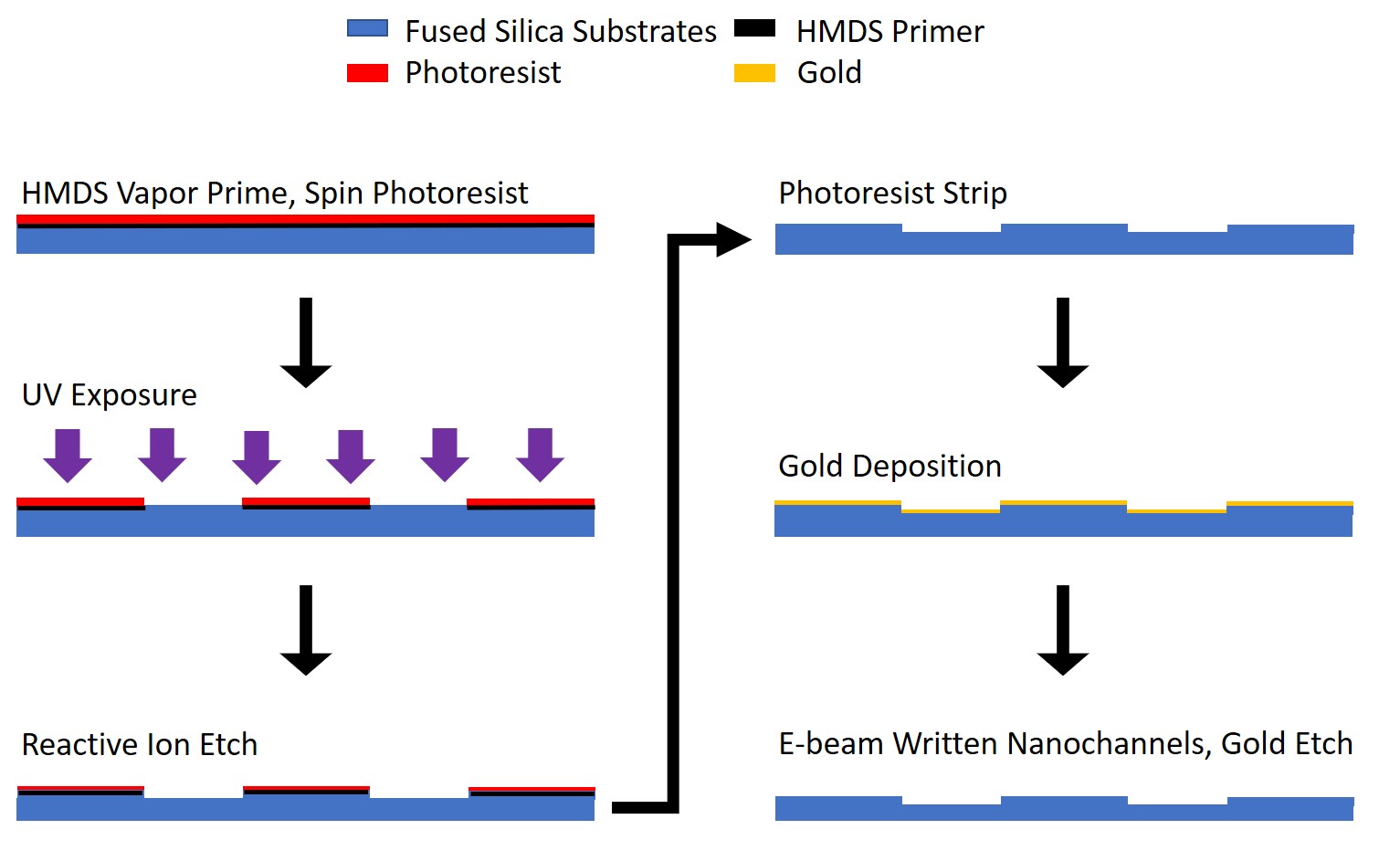}
\caption{\small Cross-sectional nanochannel process flow. Directions of steps indicated by arrows. First, start with an HMDS vapor primed fused-silica wafer coated in 1.8 $\mu$m AZ1518 photoresist. Next, perform UV photolithography exposure with indicated photomask. RIE exposed pattern to etch it into the substrate. Strip wafer in microposit remover 1165 photoresist stripper at 70$\degree$ C. This allows microchannels of width, $w=50 \mu$m, and depth, $d=1.5 \mu$m to be present on substrate.  Prepare wafer with 15 nm gold for E-beam writing. Gold etch after the nanochannel pattern is written.}
\label{Fig:Process_flow}
\end{figure*}

\subsection{Fabrication of Microchannels}
The devices were fabricated on 500 $\mu$m  thick fused-silica wafers with a diameter of $d=76.2$ mm. The wafers were cleaned in a Piranha solution 3:1 H$_2$SO$_4$:H$_2$O$_2$ for 5 minutes, followed by a DI water bath for 1 mintue. The wafers were rinsed with DI water and dried with N$_2$. Next, the wafers were placed in an isopropyl alcohol (IPA) bath for one minute, followed by a DI water bath for another minute. Afterwards, the wafers were rinsed with DI water and dried with N$_2$.\\

\begin{figure}[!h]
\centering
\includegraphics[width=\linewidth]{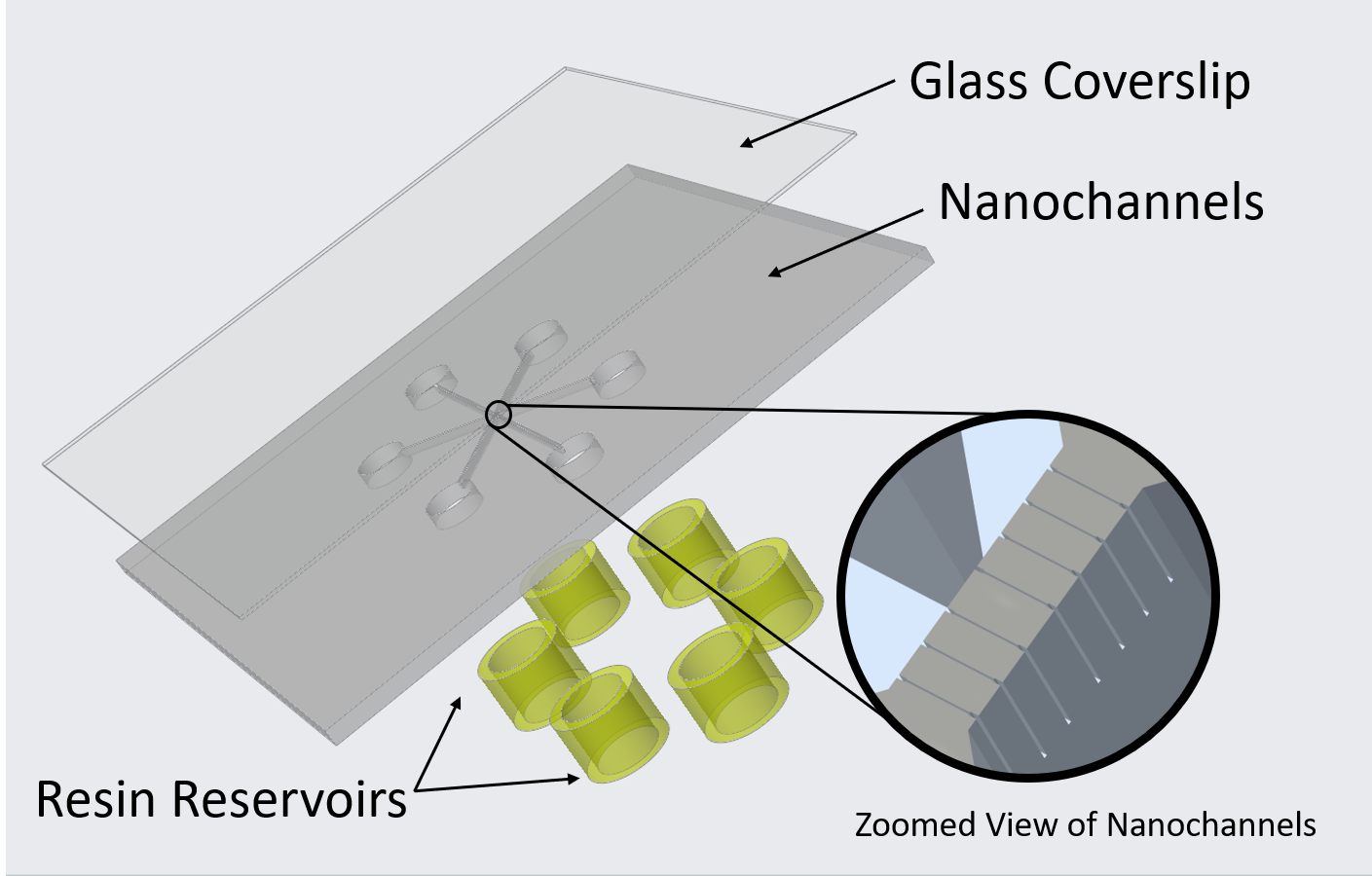}
\caption{\small Rendered CAD model of the microchannel device integrated with the nanochannels. Nanochannels are highlighted by the zoomed view insert. Imaging is done through the glass coverslip with an inverted microscope.}
\label{Fig:CAD}
\end{figure}

Next, the wafers were primed with hexamethyldisilazane (HMDS) in a Yield Engineering Systems (YES) HMDS prime oven at $148\degree$C. The procedure dehydrates the wafers and then vapor primes them with HMDS. Next, AZ1518 positive photoresist was spun on the wafers at 4000 RPM for 40 s. The wafers are baked at $110\degree$C for 2 minutes and then cooled on a cooling plate. Standard photolithography was performed for 30 s using a EVG620 mask aligner to expose the sample with the microchannel pattern shown in Figure \ref{Fig:CAD}. The wafers were manually developed in AZ300 MIF for 60 s.

Wafers were measured with a Bruker Dektak-XT profilometer to verify that the photoresist was properly removed from the patterning area. Next, the wafers were placed in an Oxford PlasmaPro 80 Reactive Ion Etch (RIE) and an O$_{2}$ descum was run for 2 minutes to remove any residual resist from the microchannel structure due to exposure and development. Next, an SiO$_2$ etch was performed for 15 minutes. The etch is a combination of CF$_4$ and O$_{2}$ gasses at 35 sscm and 3 sscm, respectively. The set forward power was 300 W and the DC bias was between 520 and 530 V. The wafers were measured with the profilometer to verify the etch depth. The photoresist was stripped for 1 hour in Dow Microposit Remover 1165, and the final depth of each channel was measured once more in the profilometer. The final depth of microchannels was around 1 $\mu$m.

In order to prep the fused-silica substrates for E-beam lithography, E-beam resist was spun on the wafers before they were coated in 15 nm of gold in an Angstrom Nexdep E-beam evaporator.

\subsection{Electron-Beam Written Nanochannels}

The pattern of nanochannels were designed to be cascading to guide MTs in(Figure 3a, 3b). Nanochannels were written between the microchannels using the Elionix ELS-G100, an electron-beam lithography system with high speed and ultra high precision thermal field emission. The ELS-G100 is capable of writing and generating patterns with a line width of 6 nm. One wafer was loaded into the FIB chamber at a time. The SEM with a beam current 500 pA was used to focus on the sample surface to determine the z-axis working distance and locate the writing area on the sample. After focusing the beams on the wafer, the pattern of nanochannels were aligned and then the wafer was ready to be exposed.

After E-beam lithography, wafers were developed in 70\% cold IPA and then inspected under microscopy to confirm that the pattern was well-written. A gold etch was then performed on the samples to remove the Au coating. Next, the wafers were placed in RIE again to etch the nanochannels with calculated parameters until the depth reached 150 nm(Figure 3d). The E-beam resist was stripped for 1 hour in Dow Microposit Remover 1165. Next, the wafers were again coated with AZ1512 photoresist to be diced into 20 mm x 20 mm devices, and transported to the lab for bonding.

\begin{figure}[!h]
\centering
\includegraphics[width=\linewidth]{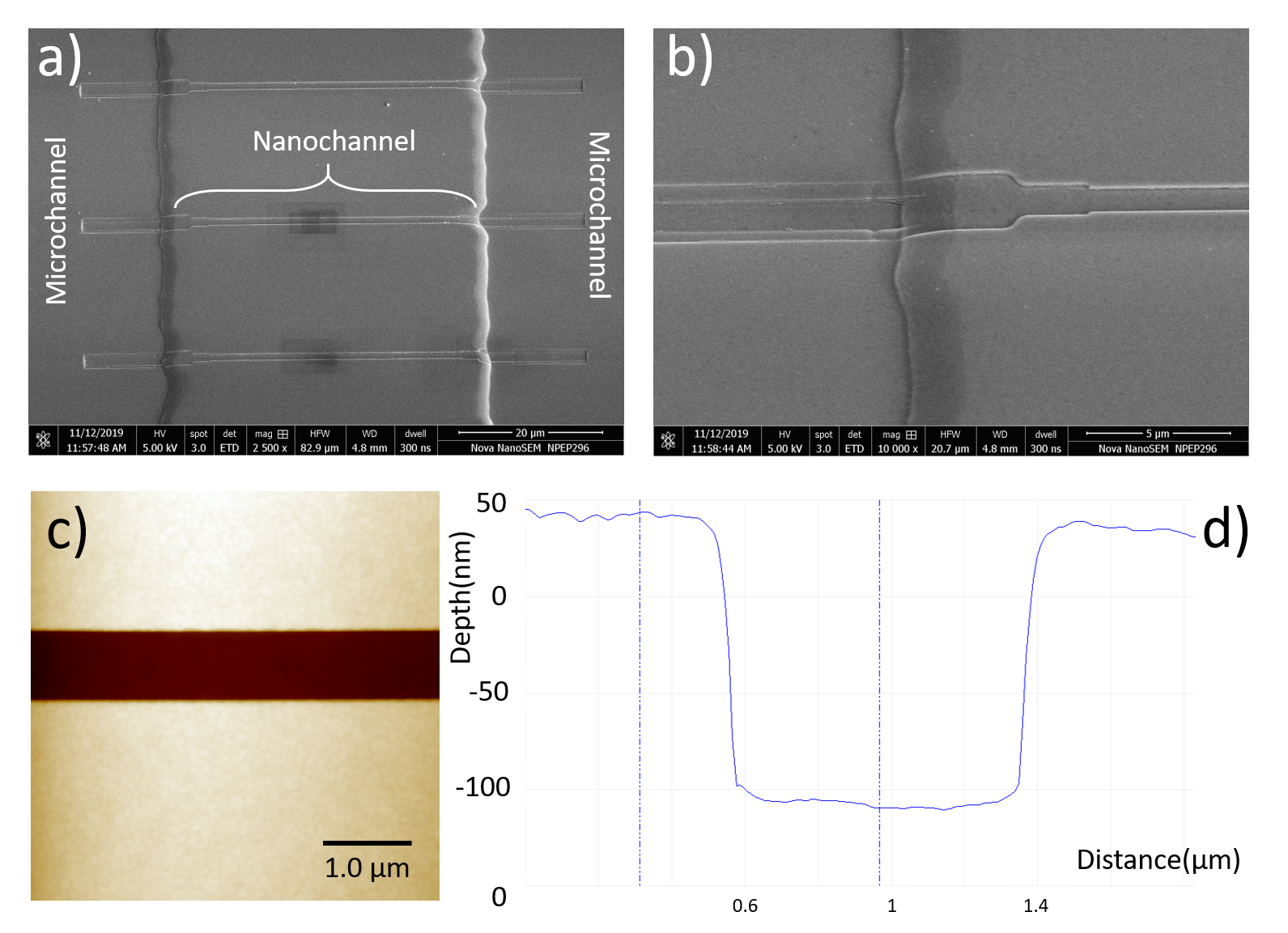}
\caption{\small a) SEM image of microchannels integrated with nanochannels. b) 10000x magnification of the cascading junction between the micron-sized and nano-sized channels. c) AFM image of one nanochannel. d) The measured depth of the nanochannel is around 150 nm.}
\label{Fig:SEM image}
\end{figure}

\subsection{Packaging Fluidic Devices}

Port holes were made at the ends of the microchannels using a BNP 220 Suction Blast Cabinet. The metal mask was made beforehand and the devices were fixed and protected by 3M Scotch Magic Tape.

To seal the devices, a modified PDMS bonding protocol was performed. To prepare for the sealing of the nanofluidic devices, it is critical that the bench top, fume hood, and laminar flow hood be as clean and dust-free as possible. They are all cleaned with 70$\%$ ethanol and again with IPA. A Ni-Lo 4 XL vacuum holder digital spin coater is cleaned, lined with fresh foil and set to 3 stages of spinning: 500 RPM for 10 s, 3000RPM for 30 s, and 5000 RPM for 20 s. The speed was reached with 500 RPM ramp speed. 

Photoresist is stripped from the diced devices with 1165 microposit remover for 15 minutes at $70\degree$C before the bonding procedure begins. The devices and 22 mm x 22 mm coverslips are cleaned by placing them in a chemically resistant coverslip rack and sonicating them in acetone for 5 minutes. Devices and coverslips are rinsed in the rack sequentially with acetone, IPA, ethanol, and DI water.

PDMS(SYLGARD 184 Silicone Elastomer Kit, Dow Corning) was mixed with its curing agent in the volume ratio of 10:1, then the mixture was placed in a desiccator for an hour to remove the bubbles. Next, 500 $\mu$l bubble-free PDMS was spun on per coverslip, coated coverslips were then moved to a pre-heated $120\degree$C hot plate, and cured for at least an hour.

Both the device(patterns face up) and coverslip(PDMS side up) were placed side by side in laminar flow hood, and treated with BD-20AC laboratory corona treater for 5 minutes. After corona treatment, the device was flipped and carefully placed on the coverslip, the bonding should happen consequently, if not, use the blunt side of a tweezers to gently tap the device on the corner. After the devices were sealed, 4 resin reservoirs were adhered to 4 ports using Loctite Epoxy instant glue, after the Epoxy was cured, the devices were ready for use.

\subsection{Flow of Microtubule into Nanochannels}

All the steps below have to be done in a humidity chamber that can be made from an empty pipette tip rack with 2 Kimwipes soaked with DI water inside. The devices should sit on the rack and the lid should be closed during waiting to prevent the device from dehydration.

The channel has to be first rinsed with PEM buffer (80 mM piperazine-N,N'-bis2-ethanesulfonic acid(PIPES), 1 mM ethylene glycol tetraacetic acid (EGTA), 1mM MgCl2, pH 6.9). 15-20 $\mu$l PEM buffer was pipetted into one port on the left microchannel, the nanochannels began filling up with liquid within 15 minutes; the excess solution from the port was pipetted out, then another 15-20 $\mu$l was placed at the port of the right microchannel. This step is to create fluid pressure for the liquid to flow in. The device was let sit in humidity chamber for another 15 minutes until all channels were filled up.

The channels then have to be passivated with BSA to prevent the adhesion of microtubules to the surface. 10 $\mu$g/ml bovine serum albumin (BSA) diluted in PEM buffer was flowed into the device. 15-20 $\mu$l BSA was pipetted into one port on the left microchannel, then waited  30 minutes for BSA to fill up the left part. The excess solution was pipetted out from the port. Another 15-20 $\mu$l was added at the port of the right microchannel. After adding the solution, the device was let sit for another 30 minutes for the channels to be filled up.

To test the quality of the seal, 1 mM fluorescein solution diluted in PEM buffer was flowed into the device. Fluorescein is sensitive to blue light, so solutions were made in with red or yellow light and foil was used to cover any containers to prevent bleaching of the sample before imaging. The nanochannels were imaged at 100x magnification with a 1.6x optovar to increase magnification. The fluorescein solution flowed into the nanochannels by placing 15-20 $\mu$l solution at one port on the left microchannel, the nanochannels began filling up with liquid within 15 minutes; the excess solution from the port was pipetted out, then another 15-20 $\mu$l was placed at the port of the right microchannel, and another 15 minutes was give for the device to sit in the humidity chamber until all channels were filled up.

Pure double cycled tubulin and tubulin labeled with Alexa594(Red) were stored in -80$^{\circ}$C in PEM buffer. To grow MTs, the tubulin was thawed on ice and grown in a GPEM buffer (1 mM guanonine triphosphate (GTP), 80 mM PIPES, 1 mM EGTA, 1mM MgCl2, pH 6.9) with 10\% dimethyl sulfoxide (DMSO). The solution was incubated on ice for 5 minutes and then incubated at 37$^{\circ}$C for one hour. The tubulin was grown at 2 mg/ml with 10\% to 50\% labeled tubulin. After incubation, warm Taxol was added to a final concentration of 10 M. The stabilized MTs were diluted 100 times in an anti-bleaching solution (250 nM glucose oxidase, 64 nM catalase, 40 mM D-glucose, 1 beta-mercaptethanol (BME), 10 $\mu$M Taxol) before flowing into fluidic devices.

We found that electrophoresys works best at guiding microtubules from the ports into the nanochannel. Electric field of 0.9 V/m was applied using a DC power supply, it took 15-30 minutes until the microtubules approached the nanochannels. 

\begin{figure}[!h]
\centering
\includegraphics[width=\linewidth]{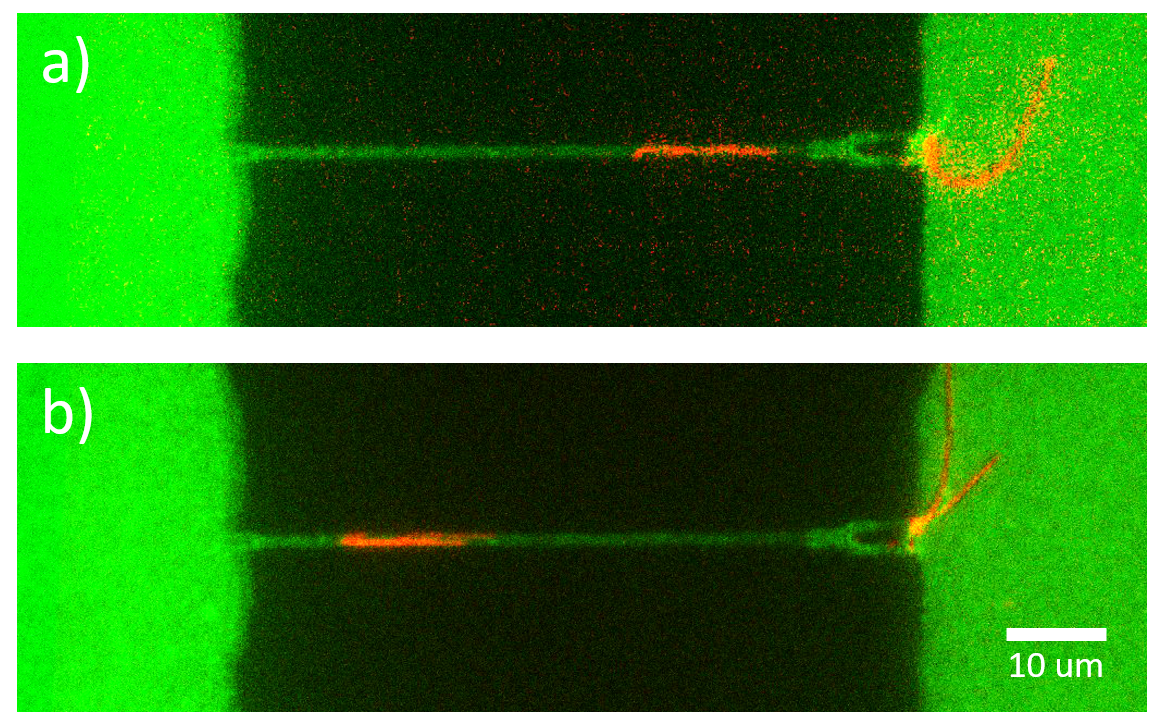}
\caption{\small Image of one fluorescence labeled microtubule(red) moving in the nanochannel filled with 1 mM fluorescein solution(green) at 100x with fluorescent microscope. b) was taken 50 millisecond after a).}
\label{Fig:nanochannels}
\end{figure}

The cleanliness of the wafer is critical for each stage of the process. Any particulates on the surface can block the fluidic channels or cause an uneven photoresist deposition. Wafers that were not cleaned properly at each step did not have precise patterning of the microchannels due to uneven photoresist, and this caused jagged edges on the microchannels. Photoresist also does not adhere well to fused-silica. Priming the wafers with HMDS was a necessary step to promote the adhesion of the photoresist to the wafers.
 
The profilometer measurements are important to verify proper photoresist development, the etch rate of the RIE, and the final depth of the microchannels. Photoresist AZ1518 is 1.8$\mu$m in depth, and the profilometer measurement after the photoresist was developed verified that the photoresist is removed from the microchannels and the wafer is set for etching. The RIE etch process etches both the fused-silica and the photoresist with different selectivity, so the etch time for a specific depth must be determined. The wafers are etched for different times, and profilimeter measurements are performed for each time to determine the etch rate. An etch time of 1 hour and 15 minutes resulted in a 1.2 $\mu$m deep microchannel.

The gold coating on the samples is necessary for SEM imaging. Fused-silica is non-conductive, and therefore charging effects due to the SEM makes it difficult to image. Charging occurs when the electrons do not have a conductive path to guide them away from the sample surface and a layer of electrons is generated. The cloud of electrons prevents clear projectile of the secondary electrons from the surface to the electron detector, resulting in poor image quality. The gold coating provides a conductive path the the electrons to move away from surface during imaging. However, long exposure to the electron beam still resulted in charging effects. 

The E-beam lithography was done on one wafer(12-21 devices depends on their size) at a time. The process is achieved by shooting a focused electron-beam at the surface of the material. The system provides a stable 1.8 nm electron-beam using high beam current at 100 kV, it is accurate and highly efficient. After exposure, the depth of nanochannels can be decided by how long they are etched in RIE.

The bonding process is robust, repeatable, and easily done outside of the clean room. A successful bonding can be seen by eye by a change in contrast between the coverslip and the fused-silica. The drawback of this method is that the sagging of PDMS material tends to partially or completely seal off the channels that do not have a high aspect ratio due to the rubber-like behavior of PDMS. However, it was noticed that if PDMS coated coverslips were let cool down after one hour heat treat and wait overnight, the chance of PDMS collapse and block the channels becomes slim.

The port holes used to be made by drilling into the ends of the microchannels using a drill press at 3600 rpm using a diamond coated drill bit of 0.5 mm diameter. The drilling of the port holes can be difficult as well as time consuming, and still result in cracking the device. Sandblasted-through portholes have cleaner edges, and the sizes of holes are well-controlled by the metal masks, also portholes of a dozen of devices can be done within a few minutes.

\section{Results and Discussion}

Previous devices for studying microtubule are microns in size. \cite{Yokokawa,Dekker}. The depth of our channels is 150nm and was chosen to keep the microtubules in the field of view of a microtscope for longer while still allowing space for dynamic instability. The width of the nanochannel is 750nm and was chosen to allow future actuation of the microtubule in the nanochannel for the purpose of exiting different types of vibrational modes in the microtubules lattice.
 Microtubules have a persistance length of 1mm making them much stiffer and longer than DNA. We found that electrophoresis was the best way to guide the microtubules into the nanochannels. We previously tried flow, and kinesin, a motor protein, but without success. For this step to be successful is very important that the bonding be very tight such that the electric field goes primary through the nanochannel. The width of the microchannel has to be large to allow the microtubules to align with the electric field prior to the entrance in the nanochannel.

In figure \ref{Fig:nanochannels} we present a microtubules flowing, under electric field through our nano channels. Prior to the flow we test the bonding and wether the channel is free by flowing flourescein (green) into the nanochannel. The microtubule (red) is flown through the nanochannel using electric field. The position of the microtubule can be controlled.
This method can be further adapted to embed electrodes on two sides of the nanochannels for sensing and actuation.

\section{Conclusion}

In conclusion, flow a microtubule through a 150 nm deep channel is reported. The channel is well characterized using SEM and AFM. The method to seal the fluidic channels allows for easy imaging with a microscope. This methodology allows for an adaptable tool to isolate a single microtubule to perform direct and quantitative measurements. The reported device is further adaptable to fabricating electrodes along the channel as an added sensor to detect and measure in the nanochannel. The small dimensions of the nanochannel are crucial for the addition of electrodes because to increase the signal to noise ratio the electrodes need to be close enough to the object of study to minimize electrical signal from the solution itself. 

The use of nanofluidic platforms has already benefited the study of DNA's unique structural properties. The adaptability and repeatability of the fabrication of the devices presented here makes it an ideal platform to open the doors to isolate and study not only DNA, but also other proteins and protein assemblies. For example, the nanochannels can be milled to an ideal height and width to isolate single actin or microtubule proteins. These biopolymers have unique polymerization and depolymerization properties that need to be quantitatively measured at the protein level to elucidate the dynamic mechanics of these structures. This platform allows for variations in size that can be optimized for the object of study. Furthermore, this platform affords the ability to isolate and manipulate macromolecules and proteins for measurements, while keeping their naturally aqueous environment.

\section{Acknowledgments}
The authors acknowledge support from the W.M. Keck Foundation. The Basuray lab acknowledges funding from the NSF grant,   1751759 ,  CAREER:"ASSURED" electrochemical platform for multiplexed detection of Cancer Biomarker Panel using Shear-Enhanced Nanoporous-Capacitive Electrodes. This research used resources of the Center for Functional Nanomaterials, which is a U.S. DOE Office of Science Facility, at Brookhaven National Laboratory under Contract No. DE-SC0012704. This work was performed in part at the Princeton Institute of the Science and Technology of Materials (PRISM) Cleanroom at Princeton University and at the Advanced Science Research Center NanoFabrication Facility of the Graduate Center at the City University of New York. The authors thank Fernando Camino, Quan Wang, and Milan Begliarbekov for their beneficial guidance and conversations.


\end{document}